# KG-Hub - Building and Exchanging Biological Knowledge Graphs

J Harry Caufield[1], Tim Putman[2], Kevin Schaper[2], Deepak R Unni[3], Harshad Hegde[1], Tiffany J Callahan[4], Luca Cappelletti[5], Sierra AT Moxon[1], Vida Ravanmehr[6], Seth Carbon[1], Lauren E Chan[7], Katherina Cortes[2], Kent A Shefchek[8], Glass Elsarboukh[2], James P Balhoff[9], Tommaso Fontana[10], Nicolas Matentzoglu[11], Richard M Bruskiewich[12], Anne E Thessen[13], Nomi L Harris[1], Monica C Munoz-Torres[2], Melissa A Haendel[2], Peter N Robinson[14], Marcin P Joachimiak[1], Christopher J Mungall[1], Justin T Reese[1*]

[1]Division of Environmental Genomics and Systems Biology, Lawrence Berkeley National Laboratory, Berkeley, CA, 94720, USA, [2]Anschutz Medical Campus, University of Colorado, Aurora, CO 80045, USA, [3]Swiss Institute of Bioinformatics, Basel, Switzerland, [4]Department of Biomedical Informatics, Columbia University Irving Medical Center, New York, NY 10032, USA, [5]Department of Computer Science, University of Milano, Milan, Italy, [6]Department of Lymphoma-Myeloma, MD Anderson Cancer Center, Houston, TX, 77030, USA, [7]College of Public Health and Human Sciences, Oregon State University, Corvallis, OR, 97331, USA, [8]Helix Genomics Inc., San Mateo, CA 94401, USA, [9]Renaissance Computing Institute, University of North Carolina, Chapel Hill, NC 27517, USA, [10]Dipartimento di Elettronica, Informazione e Bioingegneria, Politecnico di Milano, Milan, Italy, [11]Semanticly Ltd., Athens, Greece, [12]STAR Informatics, Delphinai Corporation, Sooke, BC, V9Z 0M3, Canada, [13]Department of Biomedical Informatics, University of Colorado Anschutz Medical Campus, Aurora, CO 80045, USA, [14]The Jackson Laboratory for Genomic Medicine, Farmington CT, 06032, USA

* To whom correspondence should be addressed. Email: <justinreese@lbl.gov>
Present Address: Justin Reese, Division of Environmental Genomics and Systems Biology Lawrence Berkeley National Laboratory, Berkeley, CA, 94720, USA.

## Abstract

Knowledge graphs (KGs) are a powerful approach for integrating heterogeneous data and making inferences in biology and many other domains, but a coherent solution for constructing, exchanging, and facilitating the downstream use of knowledge graphs is lacking. Here we present KG-Hub, a platform that enables standardized construction, exchange, and reuse of knowledge graphs. Features include a simple, modular extract-transform-load (ETL) pattern for producing graphs compliant with Biolink Model (a high-level data model for standardizing biological data), easy integration of any OBO (Open Biological and Biomedical Ontologies) ontology, cached downloads of upstream data sources, versioned and automatically updated builds with stable URLs, web-browsable storage of KG artifacts on cloud infrastructure, and easy reuse of transformed subgraphs across projects. Current KG-Hub projects span use cases including COVID-19 research, drug repurposing, microbial-environmental interactions, and rare disease research. KG-Hub is equipped with tooling to easily analyze and manipulate knowledge graphs. KG-Hub is also tightly integrated with graph machine learning (ML) tools which allow automated graph ML, including node embeddings and training of models for link prediction and node classification.

## Introduction

### Knowledge graphs: successes and challenges

Addressing scientific challenges such as climate change and treatment of complex or rare diseases requires integration of heterogeneous data from multiple disciplines. These datasets differ in terminology, units, granularity, and perspective, among other factors, and are difficult to combine using traditional relational databases. Knowledge graphs (KGs), which offer a more flexible and powerful way to link together heterogeneous datasets, are increasingly used to integrate data in various domains including biology, ecology, biomedicine, and personalized health (1–3). KGs represent entities (e.g. genes, diseases, phenotypes) as nodes in a graph, and relationships between these entities (e.g. gene to disease relationships) as edges between nodes. This enables the application of new types of analyses on biological data, such as network analysis and ML. The structure of the KG emphasizes the relationships between entities, which in biology is important for understanding complex systems and is often lost in more traditional data formats.

Despite the demonstrated usefulness of KGs, barriers exist that limit their effectiveness and reusability. KGs are often represented in proprietary or non-standard formats, and data models frequently differ between KGs, which greatly limits their interoperability. Biological KGs typically lack standard procedures for ID normalization (4), graph representation (multiple competing formats and knowledge representation paradigms exist) (5), data source and transformation provenance, and change tracking between KG versions (6). KG-based research also imposes computational challenges, since analyzing a typical graph with a million+ edges can require substantial memory and CPU/GPU resources (7).

Here we present a solution, KG-Hub, built to address these challenges and accelerate the construction and reuse of KGs.

### Biological and biomedical KGs

While relatively new in biology, knowledge graphs have been used in other disciplines for decades, e.g., WordNet (1985; (8)), Geonames (2005; (9)), DBpedia (2007; (10)) and the Google knowledge graph (2012; (11)). KGs are well-suited for biomedical and life science applications (12), since these fields involve highly varied types and sources of data with complex interrelationships. Graph representation of these heterogeneous relationships can generate standardized concepts and entities from multiple data sources. Previously developed biological and biomedical KG platforms address parts of this data integration process, but not the full extent of KG construction from data ingestion to reusable graph data resources.

Various strategies have been developed for constructing and applying KGs (13). In some cases, such as protein-protein interactions, the association between entities has a clear analog in a graph model: each protein is a node, while interactions between them are edges. In other contexts, the translation from data to graph requires data modeling rules. In a graph of scientific publications, for example, authorship may be represented as a connection between "author" and "paper" nodes or author names may be stored as properties of "paper" nodes, with edges representing citations. Some KGs, such as SPOKE (14) (https://spoke.ucsf.edu/), Harmonizome (15) (https://maayanlab.cloud/Harmonizome/), CROssBAR (16), and RTX-KG2 (17) unify and de-silo isolated data resources ranging from biomolecular interactions to disease risk factors and phenotypes. Other KGs are built on relationships extracted from unstructured text or from computational inference. For example, EMMAA (Ecosystem of Machine-maintained Models with Automated Analysis) assembles disease-specific graphs from published statements describing drug, gene, protein, and disease associations (18), while BioKDE (19) supports literature search by connecting related concepts. Edges may also be extracted directly from biomedical literature with rule-based or natural language processing (NLP) approaches.

## Incorporating ontologies into KGs

Ontologies provide a convenient and standardized way to add domain knowledge to a KG. There are ontologies that represent knowledge in many biological and biomedical domains; many are freely available through projects such as the OBO Foundry (20) and BioPortal (21). Adding ontologies provides valuable context to nodes and edges represented in KGs. For example, in a KG concerning dietary habits, the FOODON food ontology (22) can provide hierarchical relationships defining both raspberry jam (FOODON:03305865) and bitter orange marmalade (FOODON:03306375) as "fruit preserve or jam food product" (FOODON:00001226). The Human Phenotype Ontology (23) can be used to link individuals with a consistent set of phenotypes, e.g., abdominal pain (HP:0002027). KGs complemented with ontologies capture more of an entity's features and meaning, support complex queries (e.g., "which individuals represented in the KG ate any kind of fruit product *and* experienced abdominal pain?") and can reveal structural patterns. Including ontology data supports approaches to detect "emergent" and otherwise non-obvious associations between nodes. In the above example, incorporating diseases from the Mondo disease ontology (24) into the KG could reveal connections between digestive system disorders (MONDO:0004335) and diet.

Ontologies can be difficult to combine with each other and with instance data within KGs. Two or more resources may define concepts in conflicting ways, such as how the ChEBI ontology (25) defines penicillin (CHEBI:17334) primarily in terms of its chemical identity, while DrugCentral (26) has no specific entry for "penicillin", instead defining entries for benzylpenicillin (DrugCentral:2082) and derivatives such as ampicillin (DrugCentral:198). The potential for semantic conflict arises from merging ontology classes with instance data, e.g., the concept of "lung cancer" versus a specific lung cancer of an individual patient. Ontologies generally define a broader representation of a domain than a single set of observations is likely to cover, yet no single ontology captures the entirety of biology. A patient may experience all the symptoms consistent with a disease but not have the formal diagnosis, or a particular mutation in a given gene may not produce an identical phenotype in each case. Furthermore, our biological knowledge remains incomplete: just a subset of the phenotypes of a particular gene variant may have been observed for any individual due to multitudinous factors, such as age, genetic background, or environmental exposures. These variations present obstacles to harmonization.

## Frameworks and registries for biomedical KGs

Several frameworks exist for constructing and disseminating biological and biomedical KGs. PheKnowLator serves as a framework for standardized KG assembly with consistent formats (27), though with emphasis on adhering to Semantic Web conventions rather than a specific data model. RTX-KG2 (17) and the Knowledge Graph Exchange (KGE) Registry, both part of the NCATS Biomedical Translator project, permit researchers to assemble graphs whose contents remain intelligible and interoperable with others. Currently, creation and analysis of these graphs in NCATS Translator is closely tied to the Translator ecosystem. The Network Data Exchange (NDEx) project (28) brings together a standard and platform for exchanging graph data, including support for Cytoscape graph visualization software. The NDEx supports KG dissemination but is not generally aligned with specific data models, an element we see as crucial to reproducible graph integration and analysis.

## Graph machine learning on KGs

New knowledge can be derived from KGs using graph machine learning (graph ML). Drug repurposing is a popular use case for graph ML on KGs, as observations about drugs in therapeutic use can be integrated with basic research data from many other sources (disease to phenotype associations, protein-protein interaction data, gene functions) that provide more context for ML models. There have been notable successes in predicting new protein targets for existing drugs using graph ML over the past several years (29–31). Similarly, there have been successes in using graph ML to predict associations between genes and diseases (32), genes and phenotypes (33), or genes and functions. These efforts are supported by resources such as the Gene Ontology (GO); the inclusion of GO terms in a graph can enable function prediction using supervised graph ML (34). Graph ML on KGs has also been applied to more complex clinical challenges, such as inferring disease status given a set of symptoms and patient characteristics from electronic health record (EHR) data (35–38) or prioritizing genes by their relevance to disease phenotypes (39, 40). Recent applications have explored novel treatments for SARS-CoV-2 (41–43), predicted multiple sclerosis diagnoses (44), and provided much-needed context for protein biomarkers of fatty liver disease (45).

## KG-Hub

KG-Hub is a collection of tools and libraries for building and reusing KGs. It includes software for building interoperable KGs as well as a mechanism for sharing them. KG-Hub provides a template and documentation for creating new KG projects that follow KG-Hub design patterns and a standardized data model, Biolink Model (46). These design patterns can also be used in other KG ecosystems to preserve data provenance, provide versioned builds, adhere to data models, and incorporate ontologies.

**Materials and Methods**

## Harmonizing data sources

A wide variety of sources are used to construct knowledge graphs in KG-Hub. To unify the representation of overlapping concepts, terms, and data structures in these often disparate sources, KG-Hub uses a data model, Biolink Model, to allow cross-source interoperation. Biolink Model is an open-source data model that provides a set of hierarchical, interconnected classes (or categories) and associations that guide how entities should relate to one another. Biolink represents a wide array of biomedical entities such as gene, disease, chemical, anatomical structure, and phenotype, and establishes mappings to existing biomedical ontologies and reuses existing ontology term definition in its structure (46).

## Data downloading, transformation and graph assembly

KGs are constructed with an extract, transform, and load (ETL) process driven by the KGX and Koza toolkits and the kghub-downloader module (https://github.com/monarch-initiative/kghub-downloader). KGX (Knowledge Graph Exchange) is a KG serialization standard supported by a Python library and command line utilities that help transform data into Biolink Model compliant graphs. Koza (https://github.com/monarch-initiative/koza) is a declarative transformation framework that uses user-defined templates to both document and define the parsing and transformation of data into KGX format. KGX format. KG-Hub uses these tools to transform data sources into standalone Biolink Model compliant graphs. KG-Hub orchestrates the transformation using a *download.yaml* configuration file and a declarative transformation configuration (enabled by Koza), and a python control script. The kghub-downloader module uses the *download.yaml* file to document and manage retrieval of the data from a variety of possible sources including a local file, a URL or via queries to an Elasticsearch API (e.g. ChEMBL API). The declarative transformation configuration (as documented in the Koza framework here) and accompanying transformation script, process raw input data into KGX TSV format. KGX TSV format is a flattened serialization of the nodes and edges in a graph as tab-separated values. In addition, the declarative nature of the configuration file also serves as documentation and data instance validation for the transform (for example, it reports an error if HP:0002362 *Shuffling gait,* a phenotype, were to be typed as a Biolink:Gene). An example of this process may be found at https://github.com/Knowledge-Graph-Hub/kg-template. Koza transformations are modular, documented, reusable, and may cover multiple sources (e.g., the *ontology* transform in the *kg-template* example process is a single Python script capable of transforming any OWL ontology from the OBO Foundry). The final step in each graph's ETL process is to merge the individual transform products into one final graph. This step is handled by the KGX *merge* function and defined by a configuration file (*merge.yaml*). Once transformed, each subgraph and the merged graph are available from a centralized open repository (in Amazon Web Services, AWS), allowing users to reuse, mix, and match subgraphs, as well as share the final merged graph.

By leveraging Biolink Model, KGX, and Koza, KG-Hub graphs produced by this framework are interoperable, consistently formatted, and use the same data structures to communicate knowledge.

## Storage of graph and other data

The design of KG-Hub promotes reuse and reproduction of graph resources. Amazon Web Services (AWS) S3 is used to host files on KG-Hub, providing consistent uptime for KG-Hub. The repository is available at [https://kghub.io](https://kghub.io), and a description of the KG projects in KG-Hub is provided at [https://kghub.org](https://kghub.org).

KG-Hub provides a default template to generate modular KGs, the kg-cookiecutter ([https://github.com/Knowledge-Graph-Hub/kg-cookiecutter](https://github.com/Knowledge-Graph-Hub/kg-cookiecutter)). The kg-cookiecutter guides users through a series of prompts to customize their KG project repository, while relying on KG-Hub to specify best practices, provide basic configuration and setup, and wiring to deploy new KGs consistently. The kg-cookiecutter uses the more generic, cookiecutter framework to make its templates ([https://cookiecutter.readthedocs.io/en/stable/](https://cookiecutter.readthedocs.io/en/stable/)). The kg-cookiecutter also provides example download and merge configuration files, as well as an example transformation script that specifies the preferred location for the resulting graph and data. It stores all project materials according to KG-Hub design patterns, ensuring that the directory for each project contains the KG itself and all necessary code, data, and configurations to rebuild the graph. Users are able to customize the rebuild frequency of their KG-Hub graph using a custom Jenkins configuration. Jenkins is a continuous integration framework that helps coordinate and build systems ([https://www.jenkins.io/](https://www.jenkins.io/)), where systems are broadly defined as containers for code coupled with data.

## Querying and accessing KG-Hub resources

The availability of biomedical KGs in standard formats supports a variety of downstream use cases. KGs can be loaded into a Neo4j graph database (47) during the merge stage of the ETL process, or the KGX TSVs can be loaded directly into Neo4j using the KGX library. These Neo4j databases can then be queried using the Cypher query language (48). Some projects (e.g., KG-COVID-19) provide RDF serialization as well as Blazegraph journal files which may be queried using SPARQL. KGX TSV files can also be loaded into a tool like Cytoscape for visualization, querying, and browsing.

# Results

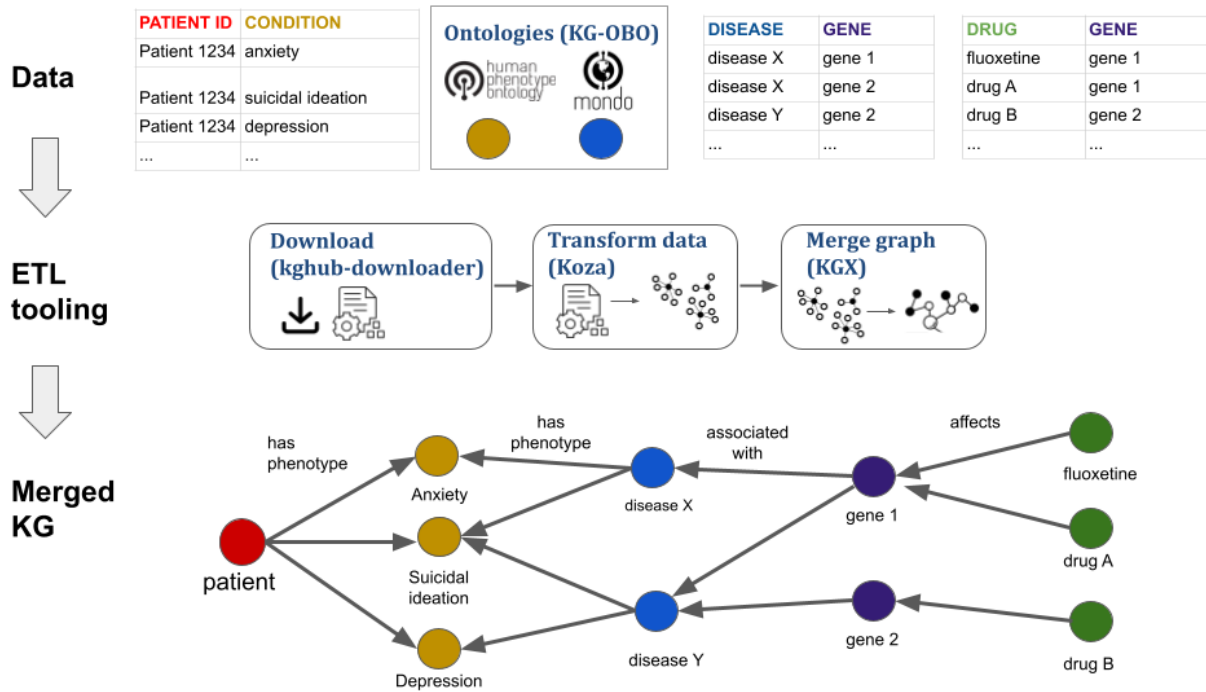

Figure 1. Integration of instance data and ontologies into knowledge graphs using KG-Hub ETL (extract, transform, and load) tooling to create new, emergent knowledge that is not present in any one data source. KG-Hub tooling comprises download (kghub-downloader), transform (Koza), and merge (KGX) components. When combined, data can provide new knowledge such as indirect relationships between patient phenotypes and drugs.

## KG-Hub design patterns

Below we describe the design patterns (reusable solutions) we developed to address commonly occurring problems in the construction and use of KGs. These design patterns can be repurposed for other KG efforts.

### Simple, modular ETL

The KG-Hub framework includes ETL Python code for transforming upstream data into KGs (https://github.com/Knowledge-Graph-Hub), with reusable, modular software for downloading data, transforming data into subgraphs, and merging subgraphs into KGs (Figure 1). See Methods for further details. In brief, the download step retrieves and saves the upstream source data, the transform step ingests and converts each upstream source data into KGX TSV format (https://github.com/biolink/kgx/blob/master/specification/kgx-format.md#kgx-format-as-tsv), and the merge step combines the subgraphs from each upstream source into a single, merged KGX TSV. The transform step in some KG projects (e.g. KG-IDG) relies on Koza

([https://github.com/monarch-initiative/koza](https://github.com/monarch-initiative/koza)), a Python package that facilitates the ingestion of data into KGX TSV format. The merge step uses KGX ([https://github.com/biolink/kgx](https://github.com/biolink/kgx)) to perform ID normalization and combine the subgraphs.

### Graph representations within KG-Hub

All graphs in KG-Hub are represented as directed, heterogeneous property graphs. Edges and nodes are typed according to a data model, edges have direction (e.g., A "affects risk for" B is distinct from B "affects risk for" A), and both nodes and edges may have one or more properties (e.g., a node may have a name or a textual description, and an edge may have a reference to a paper that provides provenance). A property graph model offers the features needed for a variety of downstream applications, such as storage in Neo4j, while also remaining sufficiently flexible to transform products to other formats (e.g., n-triples or RDF/XML).

### Biolink data model compliance

One of the identifying features of a KG built with KG-Hub is its ability to produce data harmonized according to Biolink Model. Domain knowledge in a KG that conforms to Biolink is represented using associations. An association minimally includes a subject and an object related by a Biolink Model predicate, together comprising its core triple (statement or primary assertion). A key step in KG development is identifying the concepts and relationships in Biolink Model that map to the data source being transformed. KG-Hub provides examples and guidance in selecting Biolink categories for any KG. In addition, KGX automatically assigns Biolink categories to data based on namespaces and mappings to external resources curated in Biolink Model. During data loading, nodes and edges are typed with Biolink categories and association types. With a single model that spans data sources and transformed KGs, KG-Hub facilitates analysis of KG contents in a clearly defined way, and aids interoperability of data between KG projects. For example, a KG of biolink:chemicals, biolink:proteins, and biolink:diseases can be filtered to just interactions between chemicals and proteins while retaining all subcategories. For use cases not covered by Biolink Model, users may extend the model or even create a new LinkML data model ([https://linkml.io](https://linkml.io)), then use the new model in place of Biolink in KG-Hub's graph assembly pipeline.

### Automated, self-updating, versioned builds with data provenance

In designing KG-Hub, we kept in mind that biomedical data sources are often volatile and may undergo massive updates at any time. Downloading upstream data is a frequent point of failure in bioinformatics ETL pipelines, due to issues such as changes in URLs and network instability. KG-Hub caches the most recent version of upstream data that was successfully downloaded for each data source to ensure that the ingestion can proceed even when these issues arise. KG-Hub captures data source versions and which version of its own software is used in each build and adds provenance to the data as it moves through the build process. Provenance tracking enables users and systems downstream to trace the transformation of data from its source to the resulting KG. KG-Hub also produces a permanent URL (PURL) for all artifacts in the build ensuring that downstream consumers can reproduce the results of each build even if the upstream sources have changed. Each versioned KG-Hub build, or pipeline run, is scheduled to run on a monthly basis for each KG project, using continuous integration/continuous delivery software (https://www.jenkins.io/). In addition, a new build is triggered with each update to the ETL software used in the pipeline. The build system provides error messages and guidance when volatile upstream data breaks the ETL code or fails validation.

### Easy reuse of transformed subgraphs across projects

KG-Hub is designed to allow and encourage reuse of transformed data across different projects. Each KG project produces a subgraph representing the data from each of the upstream sources that it ingests and transforms. These subgraphs are stored separately in a subdirectory (*transformed*) in the build directory for each project, with PURLs. This design allows projects to easily ingest the transformed subgraph from an upstream data source that was transformed by a different KG-Hub project, eliminating duplication of effort and encouraging alignment of data across KG projects.

For example, a new KG project that wishes to incorporate STRING protein-protein interactions in the KG could simply reuse the transformed version of this ingest from KG-COVID-19 located here:
https://kghub.io/kg-covid-19/current/transformed/STRING/
or pin to the November 2, 2022 version of the STRING ingest from KG-COVID-19 by reusing:
https://kghub.io/kg-covid-19/20221102/transformed/STRING/

### Reuse of OBO ontologies

Ontologies provide a convenient means to incorporate knowledge from domains of interest into KGs in order to contextualize instance-level data. For example, to harmonize knowledge about human diseases and disease phenotypes, one might incorporate the Mondo disease ontology (24) and the Human Phenotype Ontology (HPO) (23) into a KG. KG-OBO, a KG-Hub project that ingests and produces versioned builds of all ontologies in the OBO Foundry (20), eases the incorporation and reuse of OBO ontologies. These versioned builds can be easily incorporated into other KG-Hub projects. Individual projects can utilize the freely available ROBOT tool (http://robot.obolibrary.org/) as an interlocutor between OBO ontologies and their graph representation.

## KG-Hub projects

KG-Hub currently includes seven biomedical KG projects that integrate data pertaining to COVID-19 biology (41), drugs and drug targets, microbial phenotypes (49), and more. Each KG has a distinct set of sources, use cases, and domains. These projects exemplify KG-Hub's value as an open, general-purpose platform for exchanging biological and biomedical KGs. Figure 2 provides a summary of the types of data integrated in each project, and Table 1 provides a description of each of these projects. Projects are also documented at https://kghub.org/.

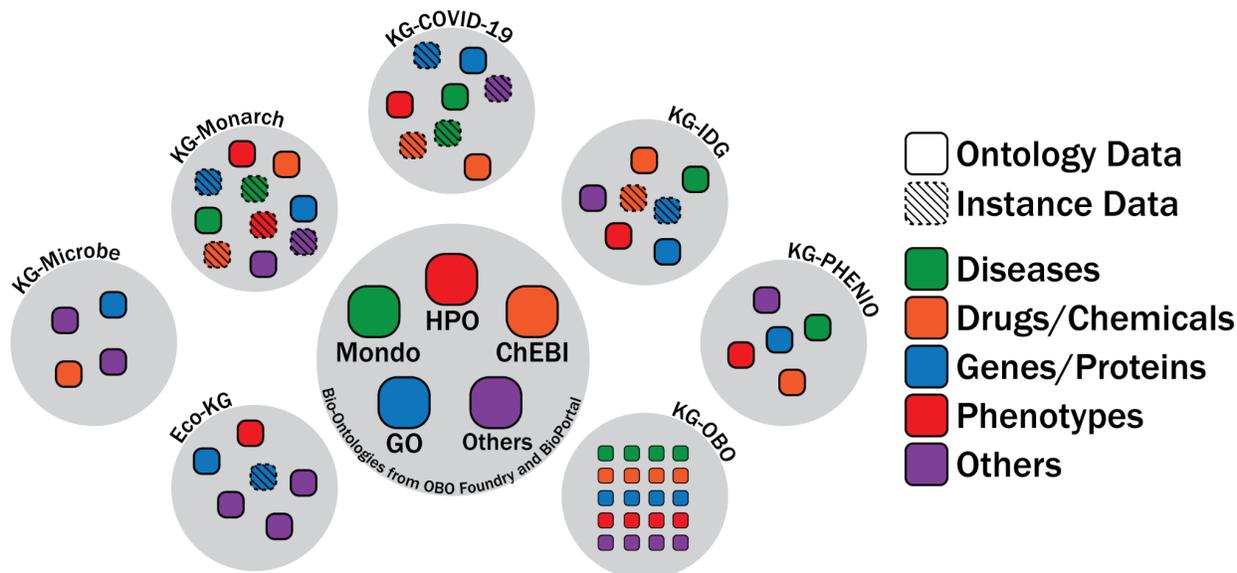

Figure 2. KG projects currently included in KG-Hub. KG-Hub currently hosts seven KG projects, which integrate disease, drug/chemical, gene/protein, phenotype and other data. Graph projects may contain both ontology and instance data. Many KG-Hub projects are constructed around a core set of ontologies related to biomedicine: GO (gene ontology/gene function), Mondo (human diseases), HPO (human disease phenotypes), ChEBI (drugs/chemicals).

| Project name | Description | Size (Nodes / Edges), in thousands | Repository URL |
|---|---|---|---|
| KG-COVID-19 | Knowledge concerning SARS-CoV-2, SARS-CoV, and MERS-CoV, including viral interactions with human proteins (41). Sourced from 10 different data sources and 4 OBO ontologies, this KG was incorporated into the N3C Enclave (50), was used in the NVBL (https://science.osti.gov/nvbl) project to provide integrated publicly available data relevant to COVID-19, and has been used to identify drugs that may affect COVID-19 outcome (51, 52). | 574 / 24,145 | https://github.com/Knowledge-Graph-Hub/kg-covid-19 |
| KG-Microbe | Data about microbial traits, environment types, carbon substrates, | 276 / 535 | https://github.com/Knowledge-Graph-Hub/kg-microbe |
| | and taxonomy. Its contents unite bacterial and archeal phenotypes across a broad range of species, supporting identification of common metabolic and environmental patterns. | | |
| KG-IDG | A graph assembled to support the Illuminating the Druggable Genome (IDG) project (https://druggablegenome.net/), with the objective of characterizing poorly-understood members of protein families that are frequently targeted by approved drugs. KG-IDG unifies structured data from 14 different sources concerning drugs, proteins, and diseases. | 560 / 4,431 | https://github.com/Knowledge-Graph-Hub/kg-idg. |
| KG-OBO | A collection of OBO Foundry (https://obofoundry.org/) ontologies transformed into JSON and graph-compatible KGX formats. 201 ontologies are currently included, many with multiple versions. | N/A | https://github.com/Knowledge-Graph-Hub/kg-obo. |
| ecoKG | Plant genes and traits, spanning 46 different species, with the objective of exploring gene, phenotype, and environment interactions. | 400 / 5,000 | https://github.com/Knowledge-Graph-Hub/eco-kg. |
| KG-Monarch | A project to integrate data relevant to human diseases, especially rare diseases (53) (https://monarchinitiative.org). This includes 12 biomedical ontologies such as HPO, Mondo, and GO, data regarding human genes, diseases, phenotypes vs. gene expression associations, as well as a range of data from many model organism databases. | 794 / 6,970 | https://github.com/monarch-initiative/monarch-ingest |
| KG-Phenio | A KG representation of the Phenomics Integrated Ontology (PHENIO) (https://github.com/monarch-initiative/phenio), a resource combining more than 20 ontologies relevant to phenotype-driven biomedical research. | 275 / 1,183 | https://github.com/Knowledge-Graph-Hub/kg-phenio |

Table 1. KG-Hub projects. For each of the seven KG-Hub projects, a description, size, and link to the project source code is provided.

## Integrated downstream tooling

KG-Hub integrates a variety of tools to facilitate functions that are frequently required downstream, such as storage, analysis and display of KG contents, conversion between different graph formats, cloud computing, and ML (Figure 3). Each module is described below.

### Visualization and querying of KG contents

A surprisingly frequent problem in previous projects was the inability to easily assess KG contents. Since KG-Hub projects use Biolink during the transform step r to type nodes and edges, KG contents can be easily inventoried and displayed. We implemented a dashboard in JavaScript to summarize the contents of KG projects, including node and edge types by source, as well as a Sankey plot showing the frequency of different pairs of node categories by source. An example of this dashboard is here: https://kghub.org/kg-hub-dashboard/.

KG-Hub projects typically emit the representation of the graph in several formats, including minimally KGX TSV format, but also RDF/ntriples and Blazegraph journal file formats. The Blazegraph journal file provides the ability to load the data into a Blazegraph instance, as well as to analyze and compare different builds of various KG projects using a tool such as Blazegraph runner (https://github.com/balhoff/blazegraph-runner).

### Integration with KGX graph utility

KGX is a tool for working with graph data that provides utilities for converting between common graph formats, extracting subgraphs, merging graphs, and loading data into graph databases (https://github.com/biolink/kgx). KG-Hub projects store graph data in KGX TSV format (https://github.com/biolink/kgx/blob/master/specification/kgx-format.md) and are therefore natively compatible with KGX. In addition to using KGX in the ETL process (in the merge step to combine subgraphs from different upstream sources), it can be used downstream to convert to other data formats, and also to load a Neo4j graph database.

### Integration with graph ML tooling

A frequent downstream use case for KGs is the application of graph ML to derive new insights. To support downstream ML use cases, we have tightly integrated with GRAPE (54), a performant graph ML package. GRAPE can import KG-Hub graphs programmatically through the API. For example, the most current KG-COVID-19 graph can be imported as follows:

```
from grape.datasets.kghub import KGCOVID19
```

```
graph = KGCOVID19()
```

This API can also retrieve a specific build for a given KG-Hub project:

```
graph = KGCOVID19(version='20210727')
```

and a list of all KG-Hub projects that are available through the API:

```
from grape.datasets import get_available_graphs_from_repository
get_available_graphs_from_repository('kghub')
```

Tight integration with GRAPE allows KG-Hub graphs to be easily downloaded, queried, and analyzed, and also allows procedures such as node embedding and graph neural networks to be applied to these KGs.

NEAT ([https://github.com/Knowledge-Graph-Hub/neat-ml](https://github.com/Knowledge-Graph-Hub/neat-ml)) is a Python package that allows graph ML algorithms to be applied to KG-Hub projects in a simple, declarative, YAML-driven way. This YAML file serves to describe the exact task, and after completion serves as provenance and documentation. An example YAML file used in the KG-IDG project is shown in Supplemental Figure 1. NEAT in turn uses GRAPE to execute the graph ML task.

### Integration with cloud computing services

All KG-Hub artifacts are stored on cloud storage (AWS S3). During the build process, index.html files are constructed to make these artifacts browsable, and these are made public using a content delivery solution (AWS CloudFront). To run ML tasks, we have integrated with a cloud computing platform, Google Cloud Platform (GCP). Within our continuous integration software configuration (Jenkins), we create new GCP instances to run the ML task, after which ML artifacts are uploaded to our browsable S3 cloud storage An automated scheduler script runs each pipeline (NEAT, etc.) on GCP and returns results to the project directory on KG-Hub.

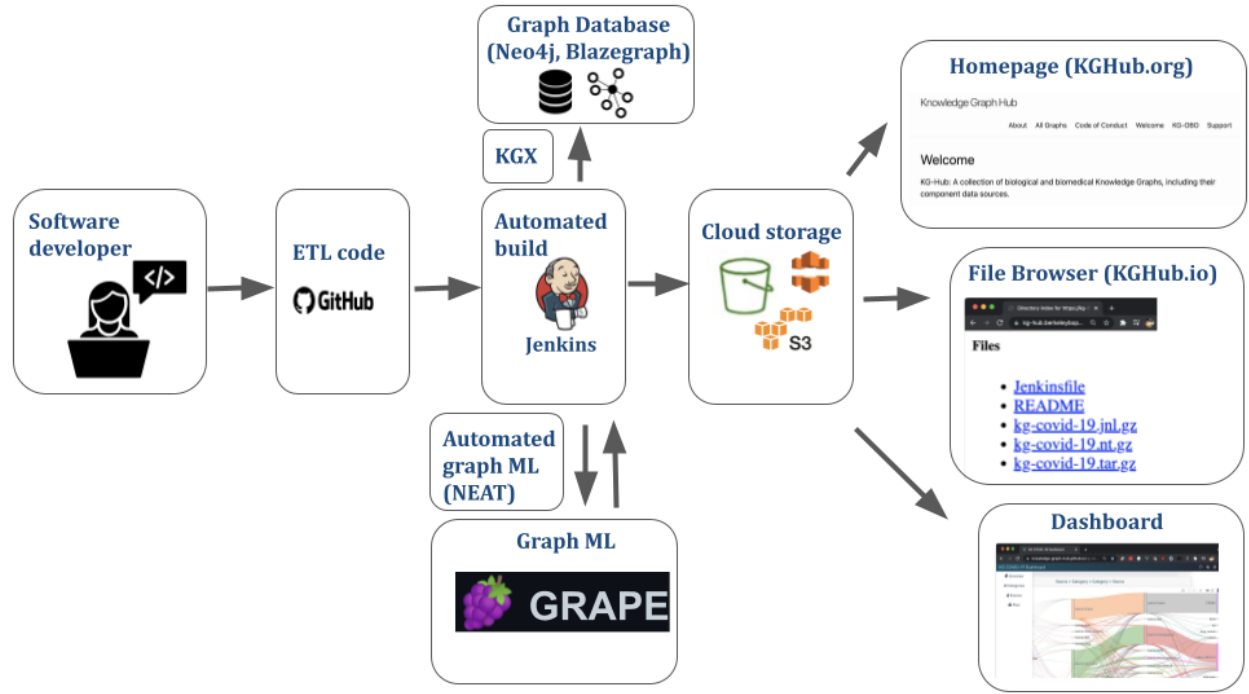

Figure 3. Schematic of tooling integrated into KG-Hub. Software developers store ETL code on GitHub. Automated builds are orchestrated on a KG-Hub server using Jenkins. Optionally, graph ML tasks can be specified for each build using NEAT yaml, and are executed using GRAPE. KGs can be directly loaded into graph databases (Neo4j, Blazegraph) using KGX, a Python library for working with graphs. Graph builds, graph ML output, provenance, and other artifacts are stored on the cloud (S3). Project summary data can be browsed on KGHub.org, and graphs and other artifacts can be browsed and downloaded on KGHub.io. A dashboard (https://kghub.org/kg-hub-dashboard/) displays detailed graph statistics for KG projects.

KG construction template

We have created a template to facilitate new KG projects (https://github.com/Knowledge-Graph-Hub/kg-cookiecutter). In addition to providing consistent project structure, the template includes example Python code for ETL functions, as well as tests and automation configurations.

Minimum requirements for new projects

The minimal necessary requirements for a new, KG-Hub-compatible project are modest by design: the project must be a biological KG and must make at least one KG build available in KGX format.

Additional guidelines serve to enhance a project's integration with the broader KG-Hub system. A project should:
- Have its own repository within the Knowledge-Graph-Hub GitHub organization or plan to move an existing repository there.
- Include ETL code and configurations that produce a KG.

- Provide a KG for public download, following semantic versioning guidelines.
- Provide the KG in additional serialization formats as needed (e.g., n-triples or JSON).

Where appropriate, each KG should model data using Biolink Model and utilize ontologies from the OBO Foundry. Each project's creator is responsible for confirming the accuracy of the datasets composing their KG and keeping track of evidence and provenance for assertions within their KG. We also strongly encourage each new project to include documentation describing the KG's intended applications, its contributors, contribution guidelines, a code of conduct, and an open license agreement.

No KG project is required to implement all of these patterns, but this enables integration with KG-Hub features. For example, automated indexing of graph components and statistics is only possible when graph components are in expected formats and predetermined locations. Data sharing limitations may also preclude following KG-Hub guidelines.

Process for creating a new KG-hub project

Assembling a KG-Hub project entails configuration, building, and analysis. Starting with the project template, a new project must be set up to retrieve data sources, transform them as needed, and merge the resulting subgraph products. The additional code necessary to perform a given transformation may be minimized by using the existing subgraphs provided on KG-Hub (e.g. DrugCentral drug-target interactions from KG-COVID-19; ontologies such as HPO, GO, or ChEBI from KG-OBO, or even entire other KG projects). After merging transformed sources with KGX, the resulting graph may then be examined and analyzed further with GRAPE and NEAT, as described above.

**Discussion**

KG-Hub addresses many hurdles involved in assembling, sharing, and using KGs. While KGs provide an elegant method to integrate biological data, work involving KGs remains challenging for a variety of technical and sociological reasons. Analyses of KGs are difficult to reproduce without versioned builds of KGs (as provided by KG-Hub), and interpretation of these analyses are often confounded by lack of provenance and lack of access to upstream data. Integrating data across multiple KGs is challenging due to incompatibilities in format, schema, and data representation. KG-Hub helps to lower these barriers.

KG-Hub could be considered the analog of an OBO Foundry for KGs. Much of KG-Hub's philosophy echoes that of the OBO Foundry (20), specifically the concept that data collections should follow a consistent format and be obtained from a persistent source. Beyond standardization of KGX format graph files and the Biolink data model, KG-Hub provides a blueprint and specific templates for data ingestion, transformation, and graph assembly. The unification of methods, formats, models, validation, and analysis strategies serves as an ecosystem for reproducible research that uses KGs. Demonstrations of relevant use cases are provided in the KG-Hub GitHub repository template (https://github.com/Knowledge-Graph-Hub/kg-template/tree/master/tutorials).

KG-Hub is designed to be straightforward to use and comprehensible to both seasoned KG engineers and beginning researchers. Some understanding of graph theory, ontologies, and the idiosyncrasies of various graph representations (e.g., how relationships between entities are described within triples vs. the property graph model that is used in KG-Hub) helps. For ontologies in particular, users must remain aware of the limitations of each set of axioms and terms and how they relate to a chosen domain. If chemicals are to be represented by ChEBI terms, for example, all newly-added chemical nodes should be linked to ChEBI identifiers. If the community identifies needs outside KG-Hub's current technical capabilities, we will explore further improvements. Our first point of contact for technical requests is the KG-Hub Support repository on GitHub (https://github.com/Knowledge-Graph-Hub/knowledge-graph-hub-support).

Future improvements to KG-Hub will largely depend on community needs and contributions. We will continue to include new KG projects and provide self-contained versions of other graph collections. We foresee creating further valuable assets for graph-based biological and biomedical data analysis. KG-Hub will continue to be an evolving resource for driving reproducible, standardized KG construction and reuse.

## AVAILABILITY

Artifacts (KGs, provenance, statistics) from KG-Hub projects mentioned in this paper can be obtained from the KG-Hub web site (https://kghub.io). A template project repository is available (https://github.com/Knowledge-Graph-Hub/kg-dtm-template). Source code for ongoing KG-Hub projects is available at the following GitHub repositories: KG-COVID-19 (https://github.com/Knowledge-Graph-Hub/kg-covid-19)
KG-Microbe (https://github.com/Knowledge-Graph-Hub/kg-microbe)
KG-IDG (https://github.com/Knowledge-Graph-Hub/kg-idg)
KG-OBO (https://github.com/Knowledge-Graph-Hub/kg-obo)
Eco-KG (https://github.com/Knowledge-Graph-Hub/eco-kg)
KG-Monarch (https://github.com/monarch-initiative/monarch-ingest)
KG-PHENIO (https://github.com/Knowledge-Graph-Hub/kg-phenio).

## ACKNOWLEDGEMENTS

This work was supported by the Monarch Initiative (NIH / OD #5R24OD011883); the Phenomics First Resource, a Center of Excellence in Genomic Science (NIH / NHGRI #1RM1HG010860-01); Illuminating the Druggable Genome by Knowledge Graphs (NIH / NCI #1U01CA239108-01); and BioPortal (NIH / NIGMS U24 GM143402). JHC, JR, NLH, MPJ, SC, SM were supported in part by the Director, Office of Science, Office of Basic Energy Sciences, of the U.S. Department of Energy under Contract No. DE-AC02-05CH11231. PNR was supported by NIH/NHGRI 5U24HG011449-02. AET was supported by the GenoPhenoEnvo project NSF 1940330. DU, SM, RB, NH, MMT, MH, and CM were supported by NIH/NCATS 1OT2TR003449-01, 3OT2TR003449-02S2, and OT2TR003449.

**CONFLICT OF INTEREST**

The authors declare no conflicts of interest.

**SUPPLEMENTARY DATA**

```
Target:
  target_path: output_data

GraphDataConfiguration:
  graph:
    directed: False
    node_path: ./merged-kg_nodes.tsv
    edge_path: ./merged-kg_edges.tsv
    verbose: True
    nodes_column: id
    node_list_node_types_column: category
    default_node_type: biolink:NamedThing
    sources_column: subject
    destinations_column: object
    default_edge_type: biolink:related_to

EmbeddingsConfig:
  filename: embedding.csv
  history_filename: embedding_history.json
  node_embeddings_params:
    method_name: "First-order LINE"
    iterations: 20
  tsne_filename: tsne.png

ClassifierContainer:
  classifiers:
    - classifier_id: lr_1
      classifier_name: Logistic Regression
      classifier_type: sklearn.linear_model.LogisticRegression
      edge_method: Average
      outfile: "model_lr_out"
      parameters:
        sklearn_params:
          random_state: 42
          max_iter: 100
```

```
    ApplyTrainedModelsContainer:
      models:
        - model_id: lr_1
          node_types:
            source:
              - "biolink:Protein"
            destination:
              - "biolink:Protein"
          cutoff: 0.5
          outfile: lr_protein_predictions.tsv

    Upload:
      s3_bucket: your-favorite-cloud-storage
      s3_bucket_dir: target_directory
      extra_args:
        "ACL": public-read
```

Supplemental Figure 1. An example YAML file used to produce node embeddings and edge predictions with a KG-Hub graph for use in ML analysis via the NEAT package. This configuration also instructs NEAT to upload all outputs to a directory in an AWS S3 bucket and make the objects publicly viewable.

**REFERENCES**

1. Nickel,M., Murphy,K., Tresp,V. and Gabrilovich,E. (2016) A Review of Relational Machine Learning for Knowledge Graphs. *Proc. IEEE*, **104**, 11–33.

2. Su,C., Tong,J., Zhu,Y., Cui,P. and Wang,F. (2018) Network embedding in biomedical data science. *Brief. Bioinform.*, 10.1093/bib/bby117.

3. Poelen,J.H., Simons,J.D. and Mungall,C.J. (2014) Global biotic interactions: An open infrastructure to share and analyze species-interaction datasets. *Ecol. Inform.*, **24**, 148–159.

4. Badal,V.D., Wright,D., Katsis,Y., Kim,H.-C., Swafford,A.D., Knight,R. and Hsu,C.-N. (2019) Challenges in the construction of knowledge bases for human microbiome-disease associations. *Microbiome*, **7**, 129.

5. Chaves-Fraga,D., Endris,K.M., Iglesias,E., Corcho,O. and Vidal,M.-E. (2019) What Are the Parameters that Affect the Construction of a Knowledge Graph? In *On the Move to Meaningful Internet Systems: OTM 2019 Conferences*. Springer International Publishing, pp. 695–713.

6. Issa,S., Adekunle,O., Hamdi,F., Cherfi,S.S.-S., Dumontier,M. and Zaveri,A. (undefined 2021) Knowledge Graph Completeness: A Systematic Literature Review. *IEEE Access*, **9**, 31322–31339.

7. Zeng,X., Tu,X., Liu,Y., Fu,X. and Su,Y. (2022) Toward better drug discovery with knowledge graph. *Curr. Opin. Struct. Biol.*, **72**, 114–126.

8. Miller,G.A., Beckwith,R., Fellbaum,C., Gross,D. and Miller,K.J. (1990) Introduction to WordNet: An on-line lexical database. *Int. J. Lexicogr.*, **3**, 235–244.

9. GeoNames.

10. Auer,S., Bizer,C., Kobilarov,G., Lehmann,J., Cyganiak,R. and Ives,Z. (2007) DBpedia: A Nucleus for a Web of Open Data. In *The Semantic Web*. Springer Berlin Heidelberg, pp. 722–735.

11. Singhal,A. (2012) Introducing the Knowledge Graph: things, not strings. *Google*.

12. Li,M.M., Huang,K. and Zitnik,M. (2022) Graph representation learning in biomedicine and healthcare. *Nat Biomed Eng*, **6**, 1353–1369.

13. Nicholson,D.N. and Greene,C.S. (2020) Constructing knowledge graphs and their biomedical applications. *Comput. Struct. Biotechnol. J.*, **18**, 1414–1428.

14. Nelson,C.A., Butte,A.J. and Baranzini,S.E. (2019) Integrating biomedical research and electronic health records to create knowledge-based biologically meaningful machine-readable embeddings. *Nat. Commun.*, **10**, 3045.

15. Rouillard,A.D., Gundersen,G.W., Fernandez,N.F., Wang,Z., Monteiro,C.D., McDermott,M.G. and Ma'ayan,A. (2016) The harmonizome: a collection of processed datasets gathered to serve and mine knowledge about genes and proteins. *Database*, **2016**, baw100.

16. Doğan,T., Atas,H., Joshi,V., Atakan,A., Rifaioglu,A.S., Nalbat,E., Nightingale,A., Saidi,R., Volynkin,V., Zellner,H., *et al.* (2021) CROssBAR: comprehensive resource of biomedical relations with knowledge graph representations. *Nucleic Acids Res.*, **49**, e96.

17. Wood,E.C., Glen,A.K., Kvarfordt,L.G., Womack,F., Acevedo,L., Yoon,T.S., Ma,C., Flores,V., Sinha,M., Roach,J.C., *et al.* (2021) RTX-KG2: a system for building a semantically standardized knowledge graph for translational biomedicine. *bioRxiv*, 10.1101/2021.10.17.464747.

18. INDRALAB EMMAA (Ecosystem of Machine-maintained Models with Automated Assembly).

19. Pang,X., Bou-Dargham,M.J., Liu,Y., Cui,Z., Sha,L., Zhao,T. and Zhang,J. (2018) Abstract 2247: Accelerating cancer research using big data with BioKDE platform. *Cancer Res.*, **78**, 2247–2247.

20. Jackson,R., Matentzoglu,N., Overton,J.A., Vita,R., Balhoff,J.P., Buttigieg,P.L., Carbon,S., Courtot,M., Diehl,A.D., Dooley,D.M., *et al.* (2021) OBO Foundry in 2021: operationalizing open data principles to evaluate ontologies. *Database* , **2021**.

21. Whetzel,P.L., Noy,N.F., Shah,N.H., Alexander,P.R., Nyulas,C., Tudorache,T. and Musen,M.A. (2011) BioPortal: enhanced functionality via new Web services from the National Center for Biomedical Ontology to access and use ontologies in software applications. *Nucleic Acids Res.*, **39**, W541–W545.

22. Dooley,D.M., Griffiths,E.J., Gosal,G.S., Buttigieg,P.L., Hoehndorf,R., Lange,M.C., Schriml,L.M., Brinkman,F.S.L. and Hsiao,W.W.L. (2018) FoodOn: a harmonized food ontology to increase global food traceability, quality control and data integration. *NPJ Sci Food*, **2**, 23.

23. Köhler,S., Gargano,M., Matentzoglu,N., Carmody,L.C., Lewis-Smith,D., Vasilevsky,N.A., Danis,D., Balagura,G., Baynam,G., Brower,A.M., *et al.* (2021) The Human Phenotype Ontology in 2021. *Nucleic Acids Res.*, **49**, D1207–D1217.

24. Vasilevsky, N., Essaid, S., Matentzoglu, N., et al. Mondo Disease Ontology: harmonizing disease concepts across the world. In.p. Vol. 2807.

25. Hastings,J., Owen,G., Dekker,A., Ennis,M., Kale,N., Muthukrishnan,V., Turner,S., Swainston,N., Mendes,P. and Steinbeck,C. (2016) ChEBI in 2016: Improved services and an expanding collection of metabolites. *Nucleic Acids Res.*, **44**, D1214–9.

26. Ursu,O., Holmes,J., Knockel,J., Bologa,C.G., Yang,J.J., Mathias,S.L., Nelson,S.J. and Oprea,T.I. (2017) DrugCentral: online drug compendium. *Nucleic Acids Res.*, **45**, D932–D939.

27. Callahan,T.J., Tripodi,I.J., Hunter,L.E. and Baumgartner,W.A. (2020) A Framework for Automated Construction of Heterogeneous Large-Scale Biomedical Knowledge Graphs. *bioRxiv*, 10.1101/2020.04.30.071407.

28. Pillich,R.T., Chen,J., Rynkov,V., Welker,D. and Pratt,D. (2017) NDEx: A Community Resource for Sharing and Publishing of Biological Networks. In Wu,C.H., Arighi,C.N., Ross,K.E. (eds), *Protein Bioinformatics: From Protein Modifications and Networks to Proteomics*. Springer New York, New York, NY, pp. 271–301.

29. Sosa,D.N., Derry,A., Guo,M., Wei,E., Brinton,C. and Altman,R.B. (2020) A Literature-Based Knowledge Graph Embedding Method for Identifying Drug Repurposing Opportunities in Rare Diseases. *Pac. Symp. Biocomput.*, **25**, 463–474.

30. Smith,D.P., Oechsle,O., Rawling,M.J., Savory,E., Lacoste,A.M.B. and Richardson,P.J. (2021) Expert-Augmented Computational Drug Repurposing Identified Baricitinib as a Treatment for COVID-19. *Front. Pharmacol.*, **12**, 709856.

31. Zeng,X., Song,X., Ma,T., Pan,X., Zhou,Y., Hou,Y., Zhang,Z., Li,K., Karypis,G. and Cheng,F. (2020) Repurpose Open Data to Discover Therapeutics for COVID-19 Using Deep Learning. *J. Proteome Res.*, **19**, 4624–4636.

32. Hu,J., Lepore,R., Dobson,R.J.B., Al-Chalabi,A., M Bean,D. and Iacoangeli,A. (2021) DGLinker: flexible knowledge-graph prediction of disease-gene associations. *Nucleic Acids Res.*, **49**, W153–W161.

33. Mungall,C.J., McMurry,J.A., Köhler,S., Balhoff,J.P., Borromeo,C., Brush,M., Carbon,S., Conlin,T., Dunn,N., Engelstad,M., *et al.* (2017) The Monarch Initiative: an integrative data and analytic platform connecting phenotypes to genotypes across species. *Nucleic Acids Res.*, **45**, D712–D722.

34. Glass,K. and Girvan,M. (2015) Finding New Order in Biological Functions from the Network Structure of Gene Annotations. *PLoS Comput. Biol.*, **11**, e1004565.

35. Choi,E., Bahadori,M.T., Song,L., Stewart,W.F. and Sun,J. (2016) GRAM: Graph-based Attention Model for Healthcare Representation Learning. *arXiv [cs.LG]*.

36. Rotmensch,M., Halpern,Y., Tlimat,A., Horng,S. and Sontag,D. (2017) Learning a Health Knowledge Graph from Electronic Medical Records. *Sci. Rep.*, **7**, 5994.

37. Ma,F., You,Q., Xiao,H., Chitta,R., Zhou,J. and Gao,J. (2018) KAME: Knowledge-based Attention Model for Diagnosis Prediction in Healthcare. In *Proceedings of the 27th ACM International Conference on Information and Knowledge Management - CIKM '18*. ACM Press, New York, New York, USA, pp. 743–752.

38. Shang,Y., Tian,Y., Zhou,M., Zhou,T., Lyu,K., Wang,Z., Xin,R., Liang,T., Zhu,S. and Li,J. (2021) EHR-Oriented Knowledge Graph System: Toward Efficient Utilization of Non-Used Information Buried in Routine Clinical Practice. *IEEE J Biomed Health Inform*, **25**, 2463–2475.

39. Peng,C., Dieck,S., Schmid,A., Ahmad,A., Knaus,A., Wenzel,M., Mehnert,L., Zirn,B., Haack,T., Ossowski,S., *et al.* (2021) CADA: phenotype-driven gene prioritization based on a case-enriched knowledge graph. *NAR Genom Bioinform*, **3**, lqab078.

40. Yamaguchi,A., Shin,J.-M. and Fujiwara,T. (2021) Gene Ranking based on Paths from Phenotypes to Genes on Knowledge Graph. In *The 10th International Joint Conference on Knowledge Graphs*, IJCKG'21. Association for Computing Machinery, New York, NY, USA, pp. 131–134.

41. Reese,J.T., Unni,D., Callahan,T.J., Cappelletti,L., Ravanmehr,V., Carbon,S., Shefchek,K.A., Good,B.M., Balhoff,J.P., Fontana,T., *et al.* (2021) KG-COVID-19: A Framework to Produce Customized Knowledge Graphs for COVID-19 Response. *Patterns (N Y)*, **2**, 100155.

42. Domingo-Fernández,D., Baksi,S., Schultz,B., Gadiya,Y., Karki,R., Raschka,T., Ebeling,C., Hofmann-Apitius,M. and Kodamullil,A.T. (2021) COVID-19 Knowledge Graph: a computable, multi-modal, cause-and-effect knowledge model of COVID-19 pathophysiology. *Bioinformatics*, **37**, 1332–1334.

43. Zhang,R., Hristovski,D., Schutte,D., Kastrin,A., Fiszman,M. and Kilicoglu,H. (2021) Drug repurposing for COVID-19 via knowledge graph completion. *J. Biomed. Inform.*, **115**, 103696.

44. Nelson,C.A., Bove,R., Butte,A.J. and Baranzini,S.E. (2022) Embedding electronic health records onto a knowledge network recognizes prodromal features of multiple sclerosis and predicts diagnosis. *J. Am. Med. Inform. Assoc.*, **29**, 424–434.

45. Santos,A., Colaço,A.R., Nielsen,A.B., Niu,L., Strauss,M., Geyer,P.E., Coscia,F., Albrechtsen,N.J.W., Mundt,F., Jensen,L.J., *et al.* (2022) A knowledge graph to interpret clinical proteomics data. *Nat. Biotechnol.*, 10.1038/s41587-021-01145-6.

46. Unni,D.R., Moxon,S.A.T., Bada,M., Brush,M., Bruskiewich,R., Clemons,P., Dancik,V., Dumontier,M., Fecho,K., Glusman,G., *et al.* (2022) Biolink Model: A Universal Schema for Knowledge Graphs in Clinical, Biomedical, and Translational Science. *Clin. Transl. Sci.*, **15**,

1848–1855.

47. Neo4j Inc. (2019) Neo4j Graph Platform.

48. Francis,N., Green,A., Guagliardo,P., Libkin,L., Lindaaker,T., Marsault,V., Plantikow,S., Rydberg,M., Selmer,P. and Taylor,A. (2018) Cypher: An Evolving Query Language for Property Graphs. In *Proceedings of the 2018 International Conference on Management of Data*, SIGMOD '18. Association for Computing Machinery, New York, NY, USA, pp. 1433–1445.

49. Joachimiak,M.P., Reese,J.T., Hegde,H., Cappelletti,L., Mungall,C.J., Duncan,W.D. and Thessen,A.E. (2021) KG-Microbe: a reference knowledge-graph and platform for harmonized microbial information. *icbo2021.inf.unibz.it*.

50. Bennett,T.D., Moffitt,R.A., Hajagos,J.G., Amor,B., Anand,A., Bissell,M.M., Bradwell,K.R., Bremer,C., Byrd,J.B., Denham,A., *et al.* (2021) The National COVID Cohort Collaborative: Clinical Characterization and Early Severity Prediction. *medRxiv*, 10.1101/2021.01.12.21249511.

51. Reese,J.T., Coleman,B., Chan,L., Blau,H., Callahan,T.J., Cappelletti,L., Fontana,T., Bradwell,K.R., Harris,N.L., Casiraghi,E., *et al.* (2022) NSAID use and clinical outcomes in COVID-19 patients: a 38-center retrospective cohort study. *Virol. J.*, **19**, 84.

52. Chan,L.E., Casiraghi,E., Laraway,B., Coleman,B., Blau,H., Zaman,A., Harris,N.L., Wilkins,K., Antony,B., Gargano,M., *et al.* (2022) Metformin is associated with reduced COVID-19 severity in patients with prediabetes. *Diabetes Res. Clin. Pract.*, **194**, 110157.

53. Shefchek,K.A., Harris,N.L., Gargano,M., Matentzoglu,N., Unni,D., Brush,M., Keith,D., Conlin,T., Vasilevsky,N., Zhang,X.A., *et al.* (2020) The Monarch Initiative in 2019: an integrative data and analytic platform connecting phenotypes to genotypes across species. *Nucleic Acids Res.*, **48**, D704–D715.

54. Cappelletti,L., Fontana,T., Casiraghi,E., Ravanmehr,V., Callahan,T.J., Joachimiak,M.P., Mungall,C.J., Robinson,P.N., Reese,J. and Valentini,G. (2021) GRAPE: fast and scalable Graph Processing and Embedding. *arXiv [cs.LG]*.